# QUERY INTERFACE INTEGRATOR FOR DOMAIN SPECIFIC HIDDEN WEB

Sudhakar Ranjan[1], Komal K. Bhatia[2]

*Apeejay Stya University, Gurgaon[1]*

*YMCAUST, Faridabad[2]*

**ABSTRACT:**

Web is title admittance today mainly relies on search engines. A large amount of data is hidden in the databases behind the search interfaces referred to as "Hidden web", which needs to be indexed so in order to serve user's query. In this paper database and data mining techniques are used for query interface integration (QII). The query interface must resemble the look and feel of local interface as much as possible despite being automatically generated without human support.This technique keeps the related documents in the same domain so that searching of documents becomes more efficient in terms of time complexity.

## [I] INTRODUCTION:

Web is a huge hypertext information resource and increases dramatically [3].Web search engines fall into two main categories: general purpose search engines and vertical search engines. A general purpose search engine such as Google provides services for general search. However, a vertical search engine is domain specific information retrieval .vertical search engines have smaller and more manageable indexes .In order to get more accurate topic related and cohesive information, a vertical search engine might embrace data mining techniques to filter, classify, cluster data, find nontrivial knowledge and promote the search quality and relevance [4].Vertical search engine or domain specific search has been key area of research in recent years. The principle of QII is based on hierarchical mapping and DSHWC. As available in the literature [1], the crawling task of domain specific hidden web crawler (DSHWC) has been divided into five phases. Phase1 is concerned with the automatic downloading of the search interfaces. Phase2 employs Domain-specific Interface Mapper [2] that automatically identifies the semantic relationships between attributes of different search interfaces. Using these semantic mappings, the interfaces are then merged to form a Unified Search Interface (USI), which is then filled automatically. The filled USI is then submitted to the Web and the response pages thus generated are collected and analyzed i.e. single unified search interface is presented to the crawler and upon submission of a query via the interface, equivalent queries

13

are submitted to many hidden web databases via front-end query interfaces and then the results are extracted from different web-sources. The advantage of this, for example, in the "airline" domain, is to prevent querying from each airline site among large airline websites which is time consuming; the second advantage is that this integration of several interfaces into a single interface presents a simple and easy query interface that need to be filled. Then data is then downloaded from several hidden airline databases.

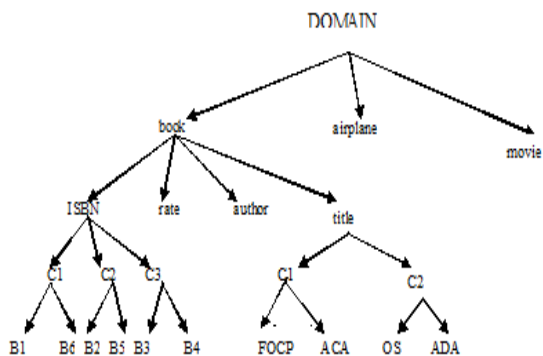

Fig. 1 Tree like structure showing organization of data among domains

In fact, it is an ordered tree of elements where each leaf correspond to a field in the interface, each non-leaf node correspond to a group or super-group of the field. In fact, the order among the sibling nodes within the tree resembles the order of fields in the interface in a left-to-right and top-to-bottom fashion. For example, <attribute 11>, <attribute 12>,<attribute 2>, <attribute3>,<attribute41>,attribute42>…

…….<attribute n> represent leaves in the interface,<attribute 1> and <attribute 2> represents non-leaf nodes and root represents the form name.

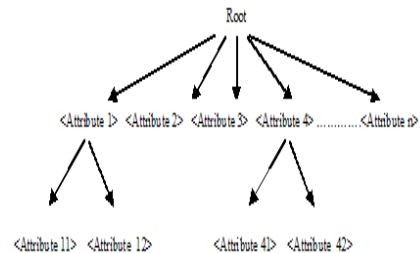

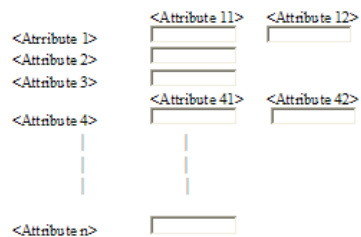

Fig. 2 Hierarchical representation of a query interface in a particular domain

## [II] MODEL AND FRAMEWORK OF A QUERY INTERFACE INTEGRATOR (QII):

Web is basically two tier architecture**.** The first tier is a web server that serves the client machine and the second tier is the client that displays that information to user. The crawler has to be able to understand and model a query interface, and the crawler has to come up with meaningful queries to issue to the query interface. In frame work of a QII based upon the concept of domain specific hidden web crawler (DSHWC) & domain specific interface mapper (DSIM).

Table 1: Brief component description

| Component | Description |
|---|---|
| DSHWC | The Domain specific hidden web crawler's goal is to automate the process of searching, viewing, filling in and submitting the search forms and analyzing the response pages. |
| Repository | A DSHWC repository stores and manages a large number of web pages. |
| Indexing Module | Indexing Module is the core part of the system. Indexes are used for searching. |
| Index DB | Store information and provide quick access to the search result. |
| Web Log | Whenever user surfs a website every web server maintains the list of actions performed/ requested by the user into a web log files. |
| Query Log | It is a collection of web query. Whenever user generated and modified query. |

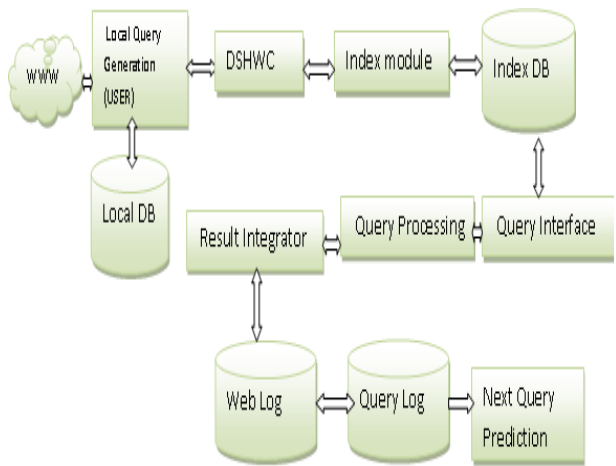

Fig.3 Work flow diagram for QII

**2.1 Local Query Generation:** A local query is generated by using the local language and data manipulation languages for each local query. In this paper used query and data manipulation language is Oracle/SQL: A local query must be sent via the network to the destination for processing by the local database, and then the result will be sent back, every local query required destination, priority and metadata information.

**2.2 Query Interface (QI):** The QI permit users to uniformly access information from local web database domain. Local query interface are organized into groups of semantically related fields. Local query interface conceptually consist of just one relation schema. Structured web databases can be queried via query forms or through web service interfaces, through query interfaces, end users are able to express their information needs by imposing selection conditions on certain attributes of interest. The matching between the attributes of a set of query interfaces in the same application domain (e.g, book, airline, movies, real state). The global interface must resemble the look and feel of local interface as much as possible despite being automatically generated without human support. Query interface organized in groups like passenger information: {Adults, Children, Infant} Option group (From, To, One way, Round Trip, Multicity, Package, Class},Status group{Leave, Return, Leave Date, Return Date} Interface group of fields organized in super groups like{ Adults, Children, Infant, From ,To, One way, Round trip, Multicity, Package, Class, Leave, Return,

Leave Date, Return Date}.When user choose One way trip then interface group of fields organized in super group appear after choosing leave status{ Adults, Children, Infant, From ,To, Leave, Leave Date} Bottom up grouping leads to a hierarchical structure for query interface.

**2.3 Query Processing**: Query processing is the procedure of transforming a high level query (SQL) into a correct and efficient plan expressed in low level languages. A set of program are required for retrievals and manipulations in the database. In query processing first phase is syntax checking .It then matches objects in the query syntax with views, relation and columns listed in the system relation. In the second phase query modification, the system validates that the user have appropriates privileges and the query does not disobey any relevant integrity constraints. One of the key challenges is to reduce the query response time is materialized view.

**2.4 Result Integrator**: Local result is sent for result integration by the network administration as a message. The results of local queries (LQ's) must be interpreted and assembled according to the global join conditions. Logically equivalent data items may be implemented differently in terms of format, scale, and encoding in different local systems. The all query schema interface integration conceptually consists of just one relation schema for domain. In this paper we used materialized view concept to improve query performance. The approach used is to pre-compute the summaries required for a query and storing it in a relation. In incremental view maintenance strategy, the changes to the base relations are used to compute the changes required to the materialized view and then update accordingly. Top view is always materialized. The new views are selected based on the benefit from the query efficiency of view materialization. The steps of this method are:

1. Materialized the top view.
2. Select additional views to materialized, one at a time, until the desired number of views is selected.

The table 2 shows the domain name and domain ID. For Query Integration we used airline domain for QI.

Table 2: Domain Name & Domain ID

Table 3: Place (R1)

Table 4: Status (R2)

Table 5: Passenger (R3)

Table 6: Materialized view

| Option | | | | | | |
|---|---|---|---|---|---|---|
| From | TO | One way | Round Trip | Multicity | Package | Class |

Table 4: Status (R2)

| Status(R2) | | | |
|---|---|---|---|
| Leave | Return | Leave Date | Return Date |
| Yes | No | 26/6/2013 | |

Table 5: Passenger (R3)

| Passenger(R3) | | |
|---|---|---|
| Adults | Children | Infant |
| 1 | - | - |

Table 6: Materialized view

| From | To | Leave | Return | Leave Date |
|---|---|---|---|---|
| Return Date | Adults | Infant | Children | Class |

The global relation consist set of tuples, set of attributes and set of real word entities. The Relation R1, R2, ….., Rn are to be integrated for a relation S. The QI does not allow the same attributes to appear multiple times, for airline domain we used three relations {Option, Status and Passenger} and one materialized view for improve query performance in term of access time, maintenance cost. The output of Query Interface containing with four attributes are {From, To, Leave and Leave Date}.

| Query Interface | Submit |
|---|---|
| From | |
| To | |
| Leave | |
| Leave Date | |
| | |

Fig.4 Query interface Integration

## [III]. RESULTS:

**3.1 Access Time Performance:** The access time performance is depending upon the size of the data amount and query integration. It is also depend upon the changing the number of keywords in the query request. The data amount of every relation is assumed as Place, Status, Passenger, Materialized view are {10K, 20K, 30K, and 60K}, and the keywords used for {Place} =1 {Place, Status} =2, {Place, Status, Passenger} =4

| Domain Nam | Domain ID |
|---|---|
| Book | 1 |
| Airline | 2 |
| Railways | 3 |
| Movie | 4 |
| Law | 5 |
| Real estates | 6 |

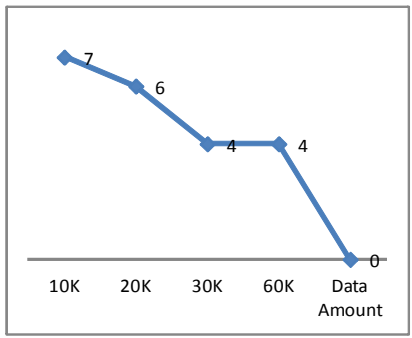

Graph 1: Access Time Performance

**3.2 Query Performance:** The costs of views Airline, Place, status, Passenger and All are 100, 80,75,70,20 respectively. A simple way of cost computation is to find the number of records in a view used to answer a query .Suppose we want to select three views. The top view is Airline. The selection of other two views is evaluated on the Airline view. Whose cost is 100.

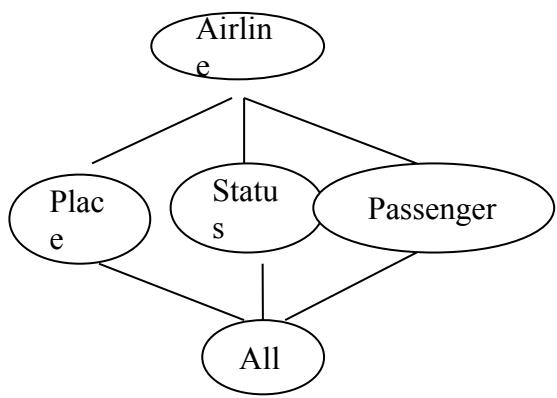

Place= (100-80)*3=60

Status= (100-75)*3=75

Passenger= (100-70)*3=90

All= (100-20)*1=80

The materialized view Passenger, All are higher than the Place, Status. So finally materialized view {Airline, Passenger, All}.

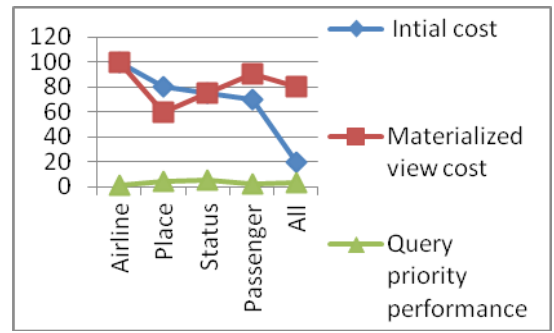

Graph 2: Materialized View Cost

**[IV] CONCLUSION:** In this paper, we propose a database and data mining techniques for query interface and integration for domain specific hidden web crawler. Specifically, we introduce the work flow diagram for query interface integration of a domain specific hidden web crawler and its core techniques. More importantly, we share our experience in building the query interface integration and searching of documents becomes more efficient.

**[V] REFERENCES:**

for Domain-Specific Search, IEEE TRANSACTIONS ON KNOWLEDGE AND DATA ENGINEERING.